\documentclass[aps,showpacs,preprintnumbers,amsmath, amssymb]{revtex4}

\usepackage{float}
\usepackage{graphics,epsfig}
\usepackage{graphicx}
\usepackage{dcolumn}
\usepackage{bm}

\begin{document}
\baselineskip=0.8 cm
\title{{\bf Quantum estimation of acceleration and temperature in open quantum system}}

\author{Zixu Zhao$^{1}$\footnote{zhao$_{-}$zixu@yeah.net}, Qiyuan Pan$^{2}$\footnote{panqiyuan@126.com} and Jiliang Jing$^{2}$\footnote{jljing@hunnu.edu.cn}}
\affiliation{$^{1}$School of Science, Xi'an University of Posts and Telecommunications, Xi'an 710121, China}
\affiliation{$^{2}$ Key Laboratory of Low Dimensional Quantum
Structures and Quantum Control of Ministry of Education, Hunan
Normal University, Changsha, Hunan 410081, China}

\vspace*{0.2cm}
\begin{abstract}
\baselineskip=0.6 cm
\begin{center}
{\bf Abstract}
\end{center}

In an open quantum system, we study the evolution of a two-level atom as a detector which interacts with given environments. For a uniformly accelerated two-level atom coupled to a massless scalar field in the Minkowski vacuum, when it evolves for a certain time, we find that there exists a peak value for the quantum Fisher information (QFI) of acceleration, which indicates that the optimal precision of estimation is achieved when choosing an appropriate acceleration $a$. QFI has different behaviors for different initial state parameters $\theta$ of the atom, displaying periodicity. However, the periodicity fades away with the evolution of time, which means that the initial state cannot affect the later stable quantum state. Furthermore, adding a boundary, we observe that the peak value of QFI increases when the atom is close to the boundary, which shows that QFI is protected by the boundary. Here, QFI fluctuates, and there may exist two peak values with a certain moment, which expands the detection range of the acceleration. Therefore, we can enhance the estimation precision of acceleration by choosing an appropriate position and acceleration $a$. The periodicity of QFI with respect to the initial state parameter $\theta$ lasts a longer time than the previous unbounded case, which indicates that the initial state is protected by the boundary. Finally, for a thermal bath with a boundary, QFI of temperature has no more than one peak value with a certain moment, which is different from QFI of acceleration with a boundary. The periodicity also lasts a longer time than unbounded case, which shows the initial quantum state of the atom is protected by the boundary for two cases.

\end{abstract}

\pacs{03.65.Vf, 03.65.Yz, 04.62.+v}
\keywords{Unruh effect; precision of parameter estimation; quantum Fisher information; quantum metrology; boundary}
\maketitle
\newpage
\vspace*{0.2cm}

\section{Introduction}

In quantum field theory, the particle content is observer dependent \cite{Fulling}. Hawking showed that quantum mechanical effects cause black holes to create and emit particles as if they were hot bodies with temperature \cite{Hawking1,Hawking2}, called Hawking radiation, which links the general relativity to quantum mechanics. Later, this procedure is applied to the Rindler coordinate system in flat spacetime \cite{Davies1}. Unruh investigated the behavior of particle detectors under acceleration \cite{Unruh}, which showed in Minkowski spacetime that the no-particle state of inertial observers (the vacuum state) corresponds to a thermal state with temperature $T_{U}=a\hbar/(2\pi c k_{B})$ for uniformly accelerated observers (here, $a$ is the observers' proper acceleration, $\hbar$ is the reduced Planck's constant, c is the speed of light, and $k_{B}$ is the Boltzmann's constant). This is usually called Fulling-Davies-Unruh effect or Unruh effect. Crispino \emph{et al.} gave a comprehensive review, which is devoted to the Unruh effect and its applications \cite{Crispino}. More recently, Lima \emph{et al.} proved that the vacuum state does induce thermalization of an accelerated extended system \cite{Lima}. Bell and Leinaas pointed out that circulating electrons in an external magnetic field can be utilized to reveal the relation between acceleration and temperature \cite{Bell}. Because $T_{U}=a\hbar/(2\pi c k_{B})$, the detection of Unruh effect would  be expected under extremely high acceleration, which is a great challenge.
In Ref. \cite{Yablonovitch}, Yablonovitch found that a nonlinear medium whose index of refraction is changing rapidly with time accelerates zero-point quantum fluctuations, and the sudden ionization of a gas or a semiconductor crystal to generate a plasma on a subpicosecond timescale can produce a reference frame accelerating at $\sim10^{20}g$ relative to an inertial frame. The detection of the Unruh effect would have important impacts in many fields \cite{Davies2,Vanzella,Gibbons,Milburn,Fuentes}. However, it is difficult to detect the Unruh effect directly.

We would like to obtain an optimal condition to detect the Unruh effect. The interpretation of quantum mechanics is probabilistic. In quantum systems, one makes quantum measurements, and the observed outcomes follow a probability distribution. As the basis of quantum metrology, estimation theory \cite{Helstrom,Holevo} presents the method to obtain the fundamental precision bounds of parameter estimation and find the optimal measurement strategies. The quantum Cram\'{e}r-Rao bound \cite{Helstrom,Cramer}, which is inversely proportional to quantum Fisher information (QFI), provides a fundamental lower bound on the covariance matrix of parameter estimation. QFI has played an important role in quantum estimation theory. It has been widely applied in the optimal quantum clock \cite{V. Buzek}, clock synchronization~\cite{N. Poli}, and entanglement detection \cite{Li} and has attracted considerable attention recently \cite{Giovannetti1,Giovannetti2,Giovannetti3,Sun,Yu,Rajabpour,Gessner,Frowis}.

In the framework of an open quantum system, we consider a uniformly accelerated two-level atom as a probe coupled to a massless scalar field in the Minkowski vacuum. Because the larger QFI corresponds the higher estimation accuracy, we will find out the condition of optimal precision of estimation by calculating the QFI of acceleration. We are curious about the result that the vacuum fluctuations are changed by the presence of a reflecting boundary. We will study how a boundary in massless scalar field influences the estimation precision of acceleration, so we calculate the QFI of acceleration for a uniformly accelerated two-level atom with a boundary. We also analyze the QFI of temperature for a static atom immersed in a thermal bath with a boundary. The QFI of the temperature is quite different from the QFI of the acceleration with a boundary.

The organization of the work is as follows. In Sec. II, we review QFI and the open quantum system. In Sec. III, for a uniformly accelerated atom coupled to a massless scalar field in the Minkowski vacuum, we analyze the estimation precision of acceleration by calculating the QFI of acceleration with and without a boundary and study the estimation precision of temperature for a static atom immersed in a thermal bath with a boundary. We will conclude in the last section of our main results.

\section{QFI and open quantum system}
In quantum metrology, QFI gives a lower bound to the mean-square error in the estimation by Cram\'{e}r-Rao inequality~\cite{Cramer,Helstrom,Holevo,Braunstein}
\begin{equation}\label{1}
\rm{Var}(X)\geq\frac{1}{N F_X}\;,
\end{equation}
where $F_X$ is the QFI of parameter $X$ and $N$ is the number of repeated measurements. Here, we calculate $F_X$ in terms of the symmetric logarithmic derivative operator as
\begin{equation}
F_X=\textrm{Tr}\,(\rho(X)L^2)\;,
\end{equation}
where $L$ is the symmetric logarithmic derivative Hermitian operator satisfying the equation $\partial_X \rho(X)=\frac{1}{2}[\rho(X)L+L\rho(X)]$. For a two-level atom system, the state is expressed in the Bloch sphere as
\begin{equation}
\rho=\frac{1}{2}(I+\bm{\omega}\cdot\bm{\sigma}
)\;,
\end{equation}
where $\bm{\omega}=(\omega_1,\omega_2,\omega_3)$ represents the Bloch vector and $\bm{\sigma}=(\sigma_1,\sigma_2,\sigma_3)$ is the Pauli matrices. The QFI of parameter $X$ therefore can be written as~\cite{Braunstein}
\begin{equation}\label{FX}
   F_X=\left\{
    \begin{array}{l}
    \overset{.}|\partial_X\bm{\omega}|^2+\frac{(\bm{\omega}\cdot\partial_X\bm{\omega})^2}{1-|\bm{\omega}|^2}\;,\;\;\,|\bm{\omega}|<1\;,  \\
    |\partial_X\bm{\omega}|^2\;,\;\;\;\;\;\;\;\;\;\;\;\;\;\;\;\;\;\;\;\;\;|\bm{\omega}|=1\;. \\
    \end{array}
    \right.
\end{equation}

For a two-level atom, the Hamiltonian of system is given by
\begin{equation}
H=H_s+H_f+H_I\;,
\end{equation}
where $H_s$, $H_f$, and $H_I$ denote Hamiltonian of the atom, the scalar field, and their interaction. Since we only pay attention to the atom and the interaction between the atom and the scalar field, their
Hamiltonians are given by
\begin{equation}
H_s={\frac{1}{2}}\hbar\omega_0\sigma_3\;,\;\;\; H_I(\tau)=(a+a_{+})\phi(t,\textbf{x})\;,
\end{equation}
where $\omega_0$ is the energy level spacing of the atom and $\sigma_3$ is the Pauli matrix; $a_+$ and $a$ are the atomic raising and lowering operators, respectively, and $\phi(t,\textbf{x})$ is the operator of the scalar field.

We take the initial total density matrix of the system $\rho_{tot}=\rho(0) \otimes |0\rangle\langle0|$; here, $\rho(0)$ is the initial reduced density matrix of the atom, and $|0\rangle$ is the vacuum state of the field. The evolution of the total density matrix $\rho_{tot}$ reads
\begin{equation}
\frac{\partial\rho_{tot}(\tau)}{\partial\tau}=-\frac{i}{\hbar}[H,\rho_{tot}(\tau)]\;,
\end{equation}
where $\tau$ is the proper time. Considering that the interaction between the atom and field is
weak, we obtain the evolution of the reduced
density matrix $\rho(\tau)$ as the Kossakowski-Lindblad form~\cite{Lindblad, pr5, Benatti2}
\begin{equation}
\frac{\partial\rho(\tau)}{\partial \tau}= -\frac{i}{\hbar}\big[H_{\rm eff},\,
\rho(\tau)\big]
 + {\cal L}[\rho(\tau)]\ ,
\end{equation}
where
\begin{equation}
{\cal L}[\rho]=\frac{1}{2} \sum_{i,j=1}^3
a_{ij}\big[2\,\sigma_j\rho\,\sigma_i-\sigma_i\sigma_j\, \rho
-\rho\,\sigma_i\sigma_j\big]\ .
\end{equation}
The coefficients of Kossakowski matrix $a_{ij}$ are
\begin{equation}
a_{ij}=A\delta_{ij}-iB
\epsilon_{ijk}\delta_{k3}-A\delta_{i3}\delta_{j3}\;,
\end{equation}
with
\begin{equation}
A=\frac{\mu^2}{4}[{\cal {G}}(\omega_0)+{\cal{G}}(-\omega_0)]\;,\;~~
B=\frac{\mu^2}{4}[{\cal {G}}(\omega_0)-{\cal{G}}(-\omega_0)]\;.\;~~
\end{equation}
Then, we introduce the two-point correlation function for the scalar field
\begin{equation}\label{G}
G^{+}(x,x')=\langle0|\phi(t,\textbf{x})\phi(t',\textbf{x}')|0 \rangle\;.
\end{equation}
The Fourier and Hilbert transformations of the field correlation function, ${\cal G}(\lambda)$ and ${\cal K}(\lambda)$, are defined as follows:
\begin{equation}
{\cal G}(\lambda)=\int_{-\infty}^{\infty} d\Delta\tau \,
e^{i{\lambda}\Delta\tau}\, G^{+}\big(\Delta\tau\big)\; ,
\quad\quad
{\cal K}(\lambda)=\frac{P}{\pi
i}\int_{-\infty}^{\infty} d\omega\ \frac{ {\cal G}(\omega)
}{\omega-\lambda} \;.
\end{equation}
By absorbing the Lamb shift term, the effective Hamiltonian $H_{\rm eff}$ is written as
\begin{equation}
H_{\rm eff}=\frac{1}{2}\hbar\Omega\sigma_3=\frac{\hbar}{2}\{\omega_0+\frac{i}{2}[{\cal
K}(-\omega_0)-{\cal K}(\omega_0)]\}\,\sigma_3\;.
\end{equation}
Assuming that the initial state of the atom is
$|\psi(0)\rangle=\cos\frac{\theta}{2}|+\rangle+e^{i\phi}\sin\frac{\theta}{2}|-\rangle$, we find that the evolution of the Bloch vector can be expressed as
\begin{align}\label{omega}
&\omega_1(\tau)=\sin\theta \cos(\Omega\tau+\phi)e^{-2 A\tau}\;,\nonumber\\
&\omega_2(\tau)=\sin\theta \sin(\Omega\tau+\phi)e^{-2 A\tau}\;,\nonumber\\
&\omega_3(\tau)=\cos\theta e^{-4 A\tau}-\frac{B}{A}(1-e^{-4 A\tau})\;.
\end{align}
\section{Parametric estimation of a two-level atom}

\subsection{Quantum estimation of acceleration without boundary}
To investigate the estimation precision of acceleration, we calculate the QFI of acceleration for a two-level atom. We adopt natural units, in which $c=\hbar=k_{B}=1$. The trajectory of a uniformly accelerated atom can be described as
\begin{eqnarray}
t{(\tau)}={\frac{1}{a}}\sinh a\tau\;,~~~
x{(\tau)}={\frac{1}{a}}\cosh a\tau\;,~~~
y{(\tau)}=y_{0}\;,~~~
z{(\tau)}=z_{0}\label{traj}\;.
\end{eqnarray}

The Wightman function of massless scalar field in the Minkowski vacuum takes the form

\begin{eqnarray}\label{wightman}
G^+(x,x')_0=-\frac{1}{4\pi^2}
           \frac{1}{(t-t'-i\epsilon)^2-(x-x')^2-(y-y')^2-(z-z')^2}\;.
\end{eqnarray}
Applying the trajectory of the atom (\ref{traj}), we obtain the field correlation function
\begin{eqnarray}
G^+(x,x')_0=-\frac{a^2}{16\pi^2}\frac{1}{\sinh^2(\frac{a\Delta\tau}{2}-i\epsilon)}\;,
\end{eqnarray}
where $\Delta\tau=\tau-\tau'$.

The Fourier transformation of the field correlation function is given by
\begin{eqnarray}\label{fourier}
{\cal G}^0(\lambda)=\frac{1}{2\pi}\frac{\lambda}{1-e^{-2\pi\lambda/a}}\;.
\end{eqnarray}

The coefficients for the Kossakowski matrix are
\begin{eqnarray}\label{ab}
\begin{aligned}
&A_0={\Gamma_0\over2}\,\coth\frac{\pi\omega_0}{a}\;,\\
&B_0={\Gamma_0\over2}\;,\\
\end{aligned}
\end{eqnarray}
with $\Gamma_0=\mu^2\omega_0/2\pi$ being the spontaneous emission rate.
In the following discussion, we use $\tau\rightarrow \tilde{\tau}\equiv{\mu^2\omega_0}\tau/{2\pi},~~~a\rightarrow \tilde{a}\equiv{a}/{\omega_0}$.  For simplicity, $\tilde{\tau}$ and $\tilde{a}$ will be written as $\tau$ and $a$ . From (\ref{FX}), (\ref{omega}) and (\ref{ab}), we can obtain the QFI of the acceleration $F_a$.

\begin{figure}[H]
\begin{centering}
\includegraphics[scale=0.55]{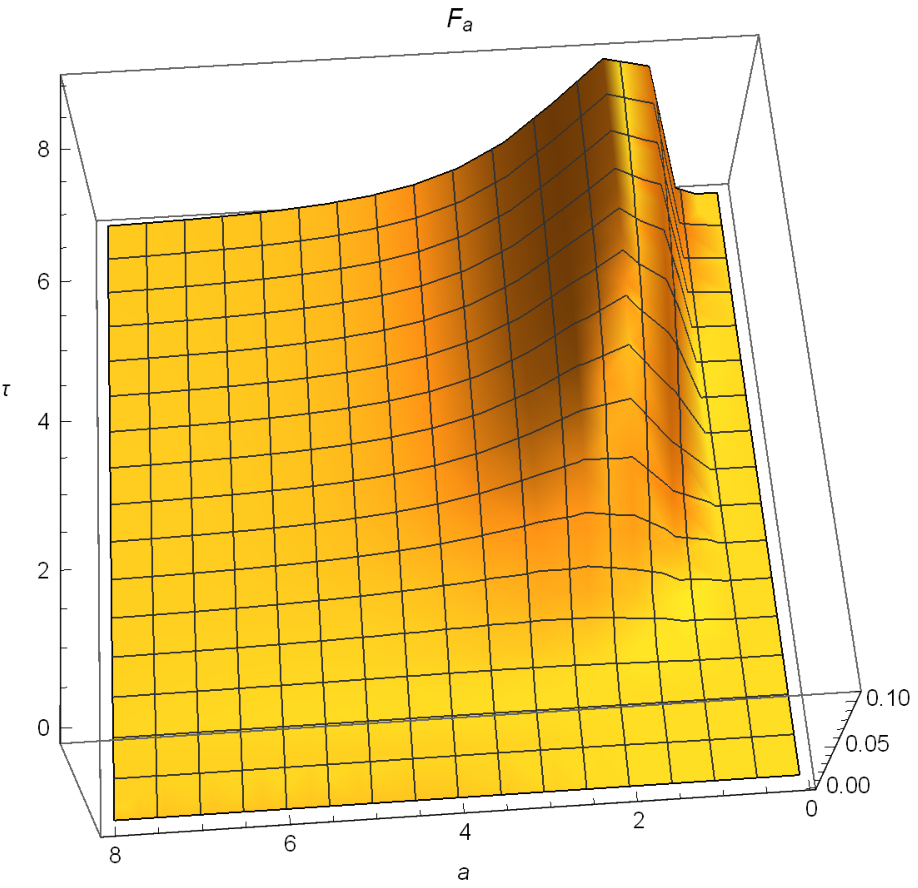}
\includegraphics[scale=0.55]{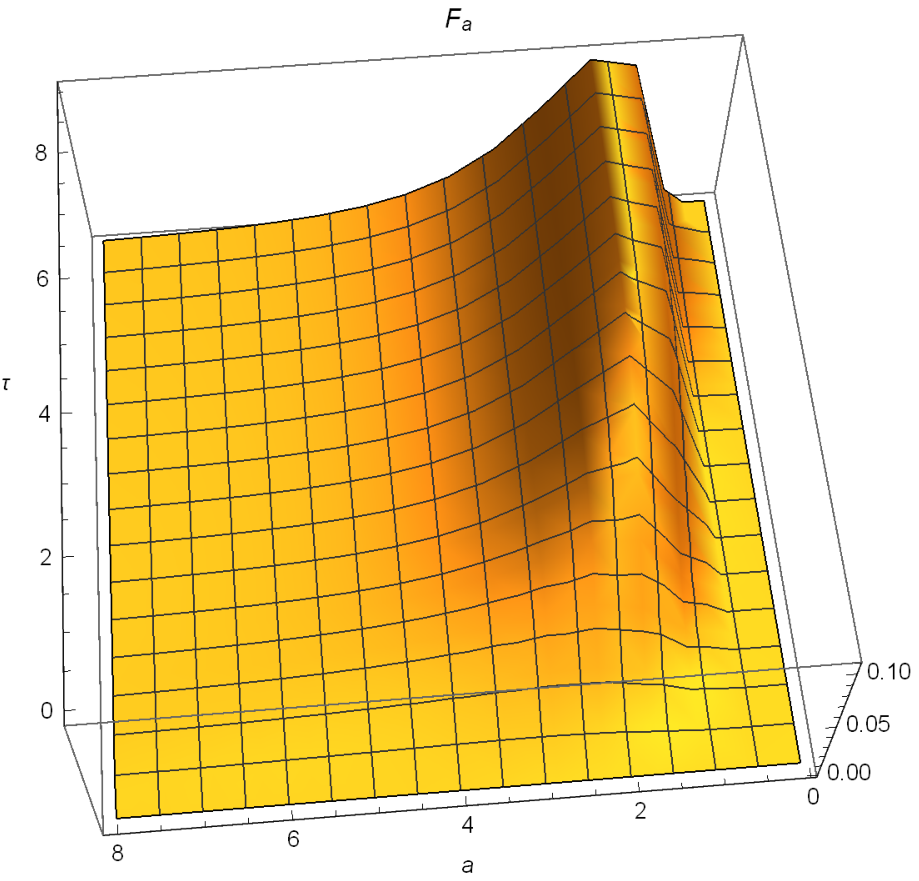}
\includegraphics[scale=0.55]{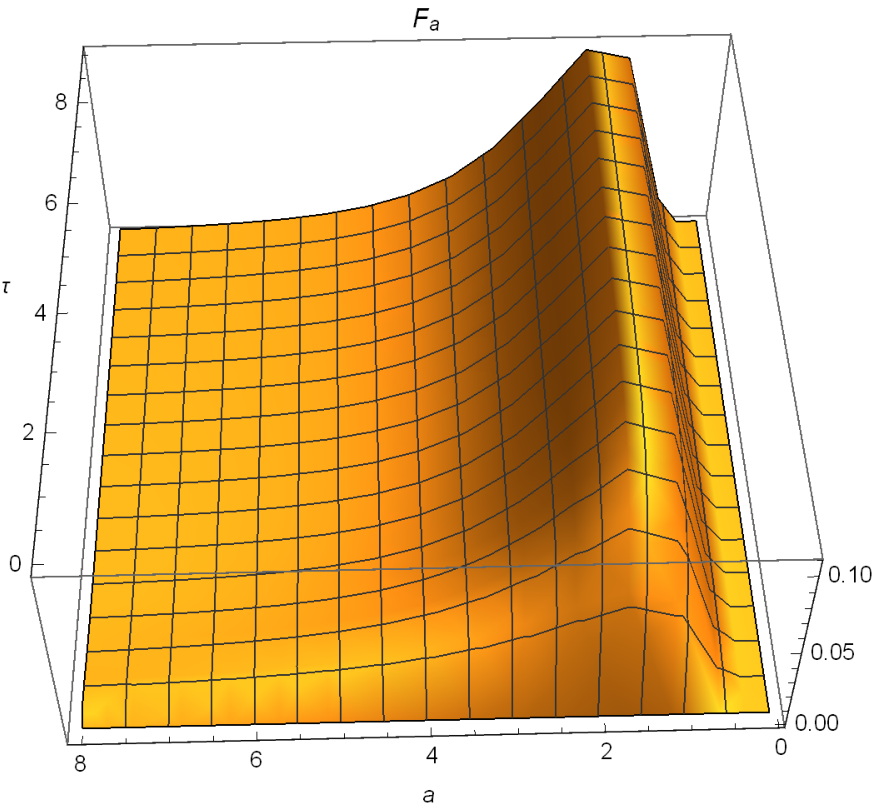}
\caption{\label{fig-Fat} The QFI of acceleration as a function of the acceleration $a$ and the evolution time $\tau$ for different initial state parameters $\theta$. We take $\theta=0$ (left panel), $\theta=\pi/2$ (middle panel), and $\theta=\pi$ (right panel). }
\end{centering}
\end{figure}

We present the QFI of acceleration $F_a$ as a function of $a$ and $\tau$ for different $\theta$ in Fig. \ref{fig-Fat}. When the uniformly accelerated detector evolves for a certain time, we observe that $F_a$  first increases and then decreases with an increase of acceleration $a$, where there exists a peak value, indicating that the optimal precision of estimation is achieved when choosing an appropriate acceleration. Compared with cases $\theta=0$ and $\theta=\pi/2$, the peak is achieved faster for $\theta=\pi$, corresponding to the ground state of the atom.

\begin{figure}[H]
\begin{centering}
\includegraphics[scale=0.56]{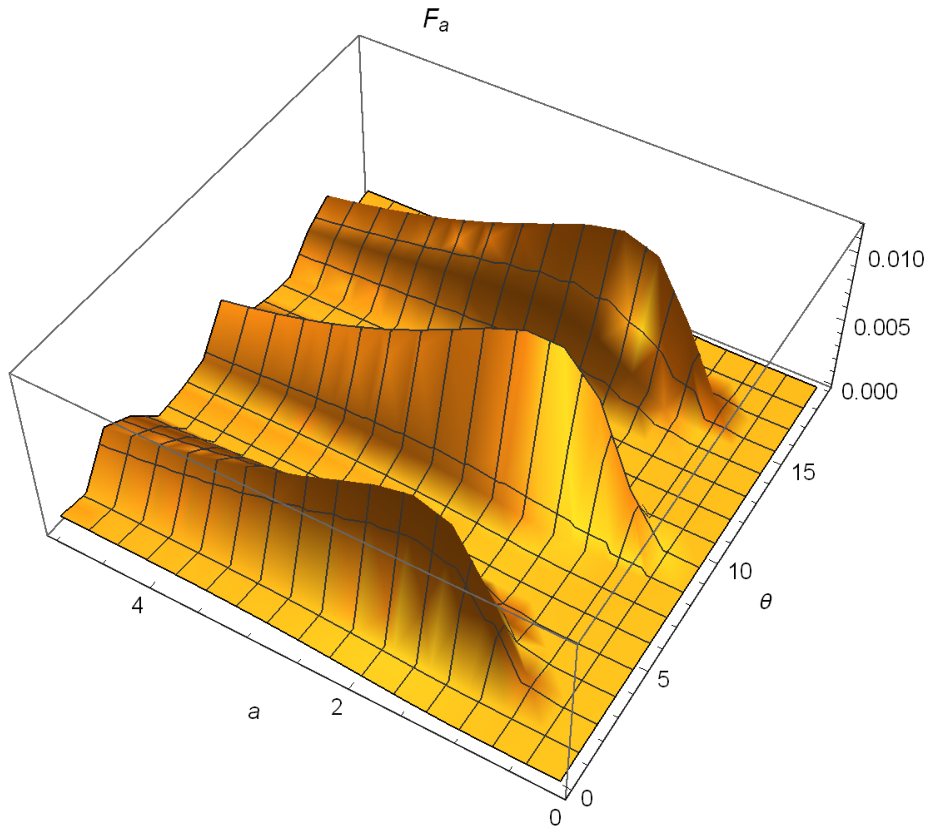}
\includegraphics[scale=0.56]{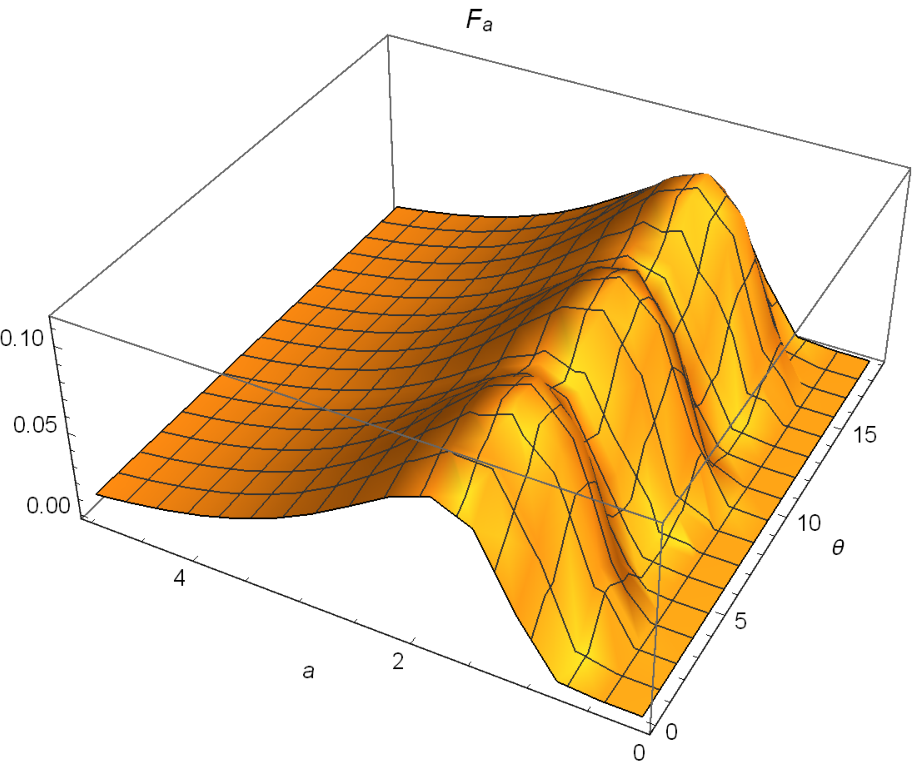}
\includegraphics[scale=0.56]{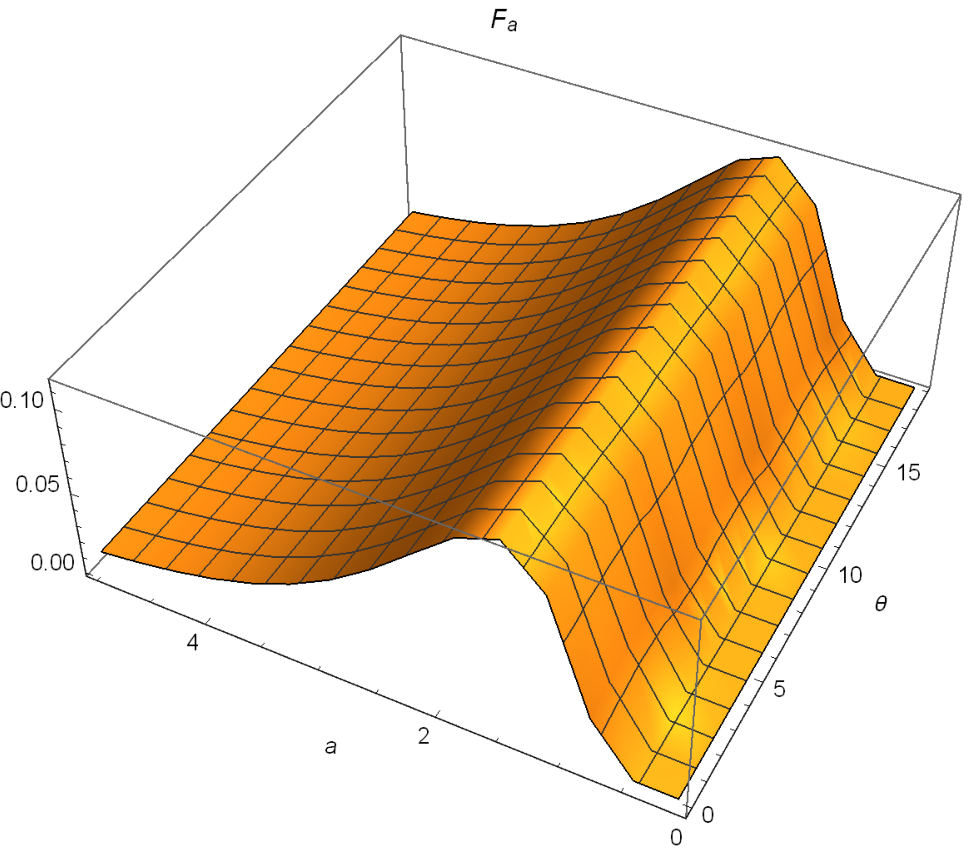}
\caption{\label{fig-Fath} The QFI of acceleration as a function of the acceleration $a$ and the initial state parameter $\theta$ for different $\tau$. We take $\tau=0.1$ (left panel), $\tau=5$ (middle panel), and $\tau=9$ (right panel). }
\end{centering}
\end{figure}

In Fig. \ref{fig-Fath}, we plot the QFI of acceleration $F_a$ as a function of $a$ and $\theta$ for $\tau=0.1$, $\tau=5$, and $\tau=9$, respectively. From the left panel (in a very short time), we see that the QFI is the periodic function of the initial state parameter $\theta$. However, the periodicity gradually fades away with the evolution of time as shown in the middle panel and the right panel, which means that the QFI can achieve the maximum for any initial state. Therefore, the initial state cannot affect the later stable quantum states.

The behavior for the QFI of temperature for the static atom immersed in a thermal bath is similar to the QFI of the acceleration due to $T_{U}=a\hbar/(2\pi c k_{B})$, although there exists a difference between the values.

\subsection{Quantum estimation of acceleration with a boundary}

We add a boundary at $z=0$ and consider an atom moving in the $x-y$ plane at a distance $z$ from the boundary. Then, the two-point function in this case is expressed as
\begin{eqnarray}
G^+(x,x')
=G^+(x,x')_0+G^+(x,x')_b\;,
\end{eqnarray}
where $G^+(x,x')_0$ is the two-point function in the unbounded case, which has already been calculated above, and
\begin{eqnarray}
G^+(x,x')_b&=&-\frac{1}{4\pi^2}
\frac{1}{(x-x')^2+(y-y')^2+(z+z')^2-(t-t'-i\epsilon)^2},\;
\end{eqnarray}
gives the correction due to the presence of the boundary.
Applying the trajectory of the atom (\ref{traj}), we obtain the field correlation function
\begin{eqnarray}
G^+(x,x')=-\frac{a^2}{16\pi^2}\left[\frac{1}{\sinh^2(\frac{a\Delta\tau}{2}-i\epsilon)}-\frac{1}{\sinh^2(\frac{a\Delta\tau}{2}-i\epsilon)-a^2 z^2}\right],
\end{eqnarray}
where $\Delta\tau=\tau-\tau'$.

The Fourier transformation of the field correlation function is given by
\begin{eqnarray}\label{fourier}
{\cal G}(\lambda)=\frac{1}{2\pi}\frac{\lambda}{1-e^{-2\pi\lambda/a}}-\frac{1}{2\pi}\frac{\lambda}{1-e^{-2\pi\lambda/a}}\frac{\sin[\frac{2\lambda}{a}\sinh^{-1}(az)]}{2z\lambda\sqrt{1+a^2 z^2}}.
\end{eqnarray}

The coefficients for the Kossakowski matrix are
\begin{eqnarray}\label{abb}
\begin{aligned}
&A_b=\frac{\mu^2 \omega_0 \coth\frac{\pi\omega_0}{a}}{8\pi}\left[1-\frac{\sin[\frac{2\lambda}{a}\sinh^{-1}(az)]}{2z\lambda\sqrt{1+a^2 z^2}}\right]     \;,\\
&B_b=\frac{\mu^2 \omega_0 }{8\pi}\left[1-\frac{\sin[\frac{2\lambda}{a}\sinh^{-1}(az)]}{2z\lambda\sqrt{1+a^2 z^2}}\right] \;.\\
\end{aligned}
\end{eqnarray}
In the following discussion, we use $\tau\rightarrow \tilde{\tau}\equiv{\mu^2\omega_0}\tau/{2\pi}$,~~~$a\rightarrow \tilde{a}\equiv{a}/{\omega_0}$, and $z\rightarrow \tilde{z}\equiv z \omega_0$. For simplicity, $\tilde{\tau}$, $\tilde{a}$ and $\tilde{z}$ will be  written as $\tau$, $a$ and $z$ . We can obtain the QFI of the acceleration $F_a$.

\begin{figure}[H]
\begin{centering}
\includegraphics[scale=0.55]{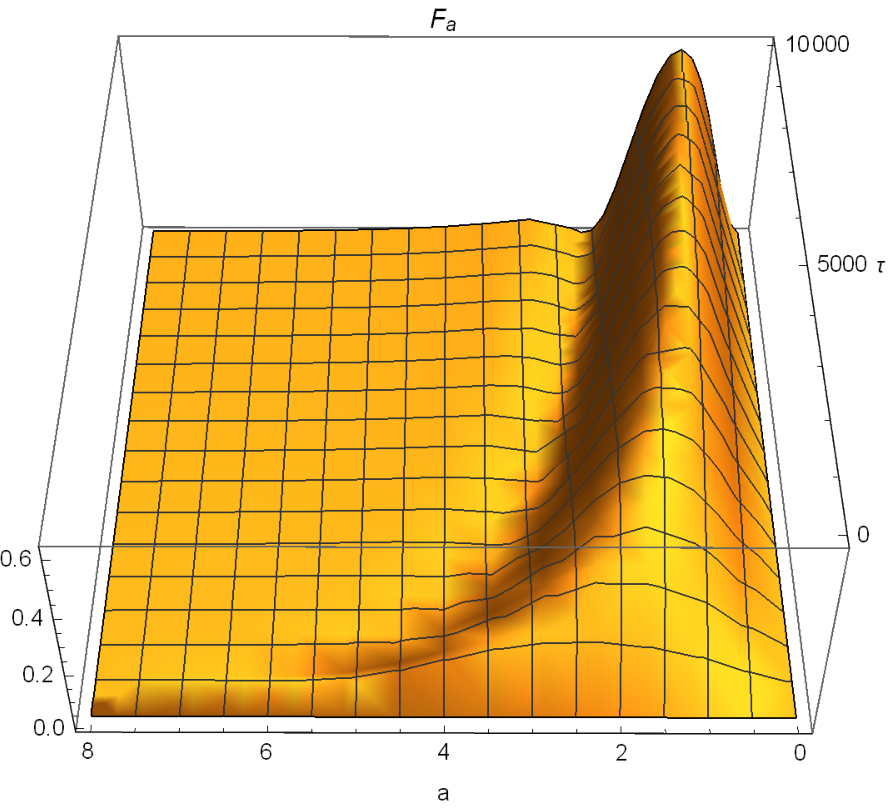}
\includegraphics[scale=0.55]{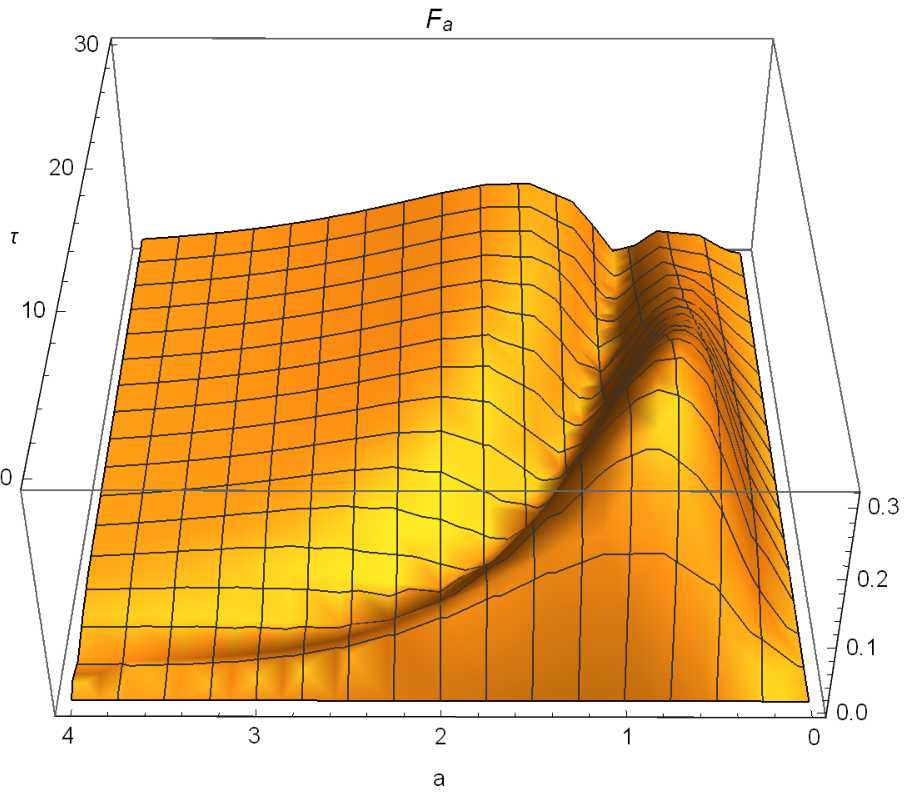}
\includegraphics[scale=0.55]{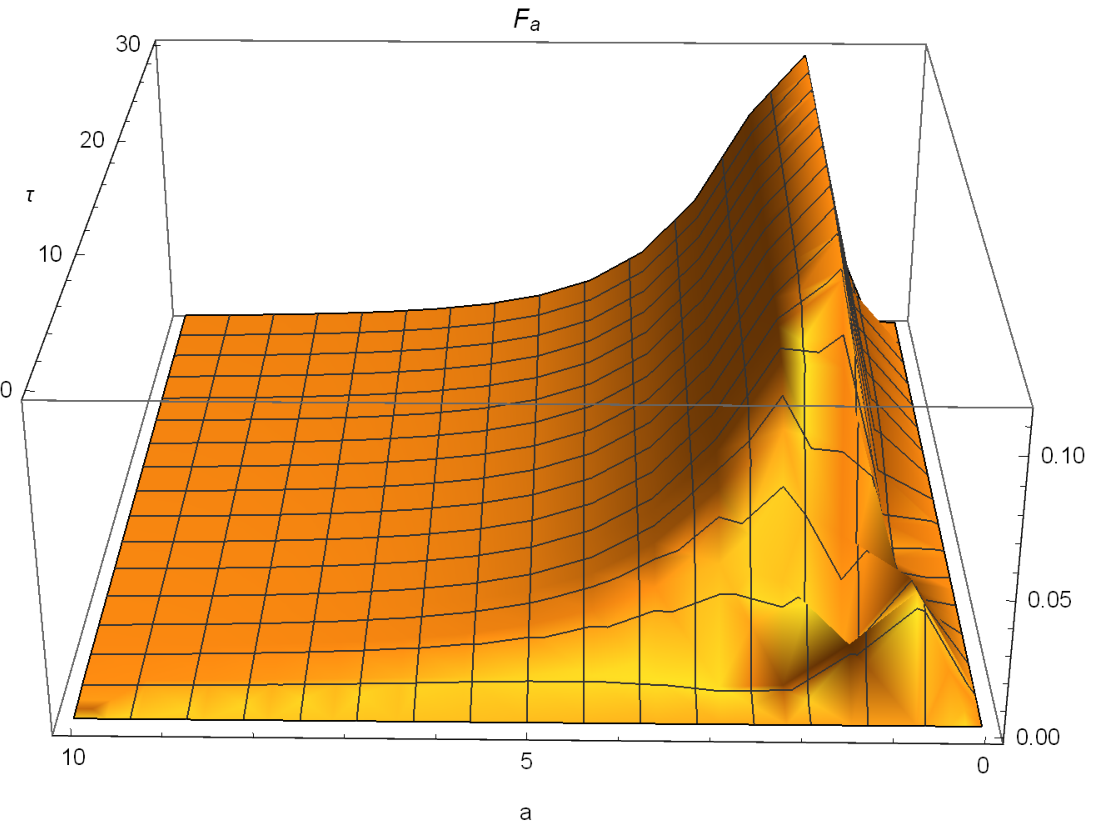}
\caption{\label{figzatF} The QFI of acceleration as a function of $a$ and $\tau$ for $\theta=0$. We take $z=0.01$ (left panel), $z=0.5$ (middle panel), and $z=1$ (right panel). }
\end{centering}
\end{figure}

We describe the QFI of acceleration as a function of $a$ and $\tau$ for $\theta=0$ in Fig. \ref{figzatF}. We find that the peak value of $F_a$ increases when the atom is close to the boundary, compared with the absence of boundary. It is obvious that the QFI is protected by the boundary, which means the estimation
precision of acceleration is enhanced by adding a boundary. Because of adding a boundary, we can see that $F_{a}$ fluctuates and there may exist two peak values with a certain moment, which expands the detection range of the acceleration.

\begin{figure}[H]
\begin{centering}
\includegraphics[scale=0.55]{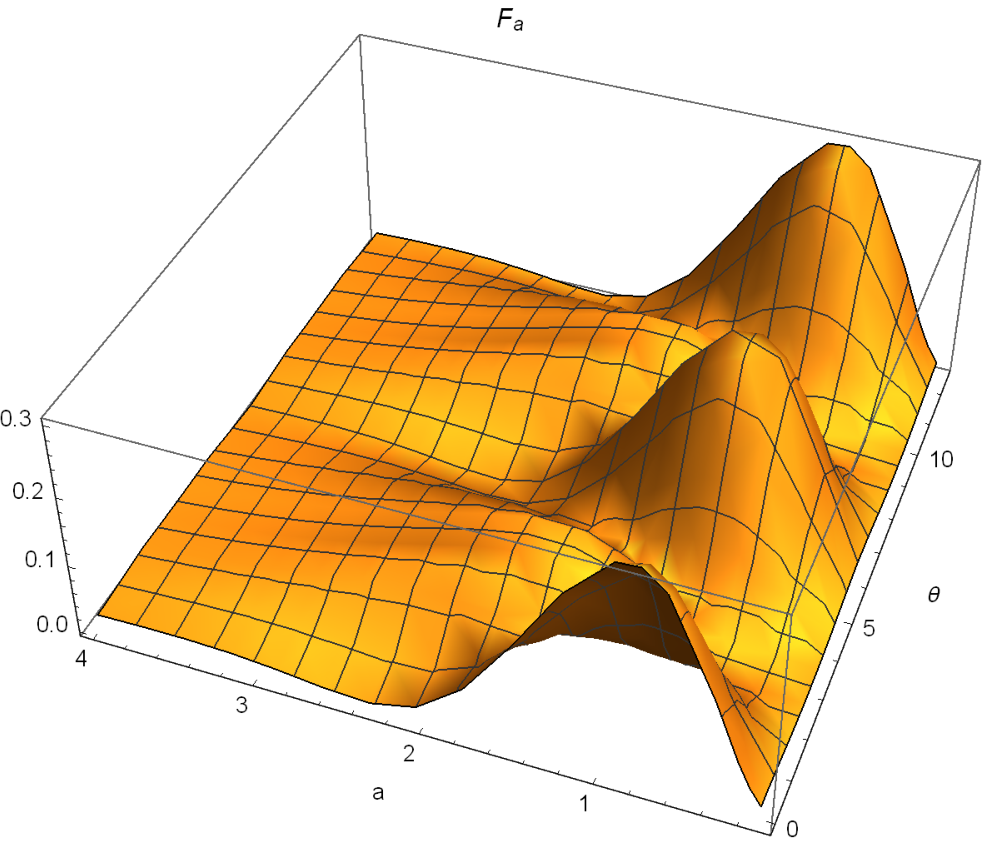}
\includegraphics[scale=0.55]{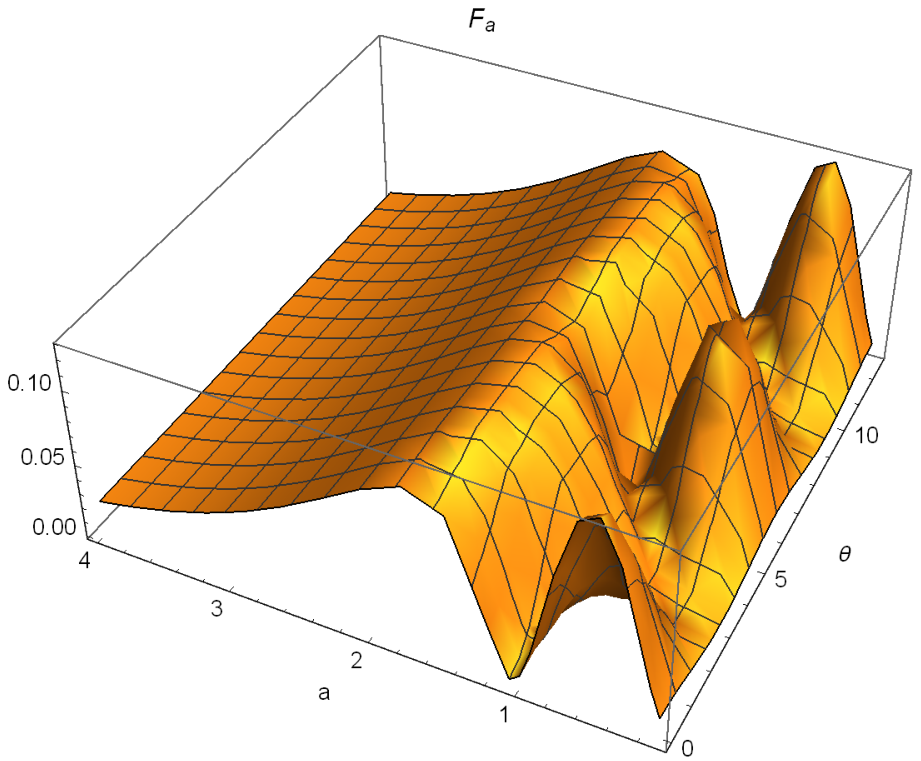}
\includegraphics[scale=0.55]{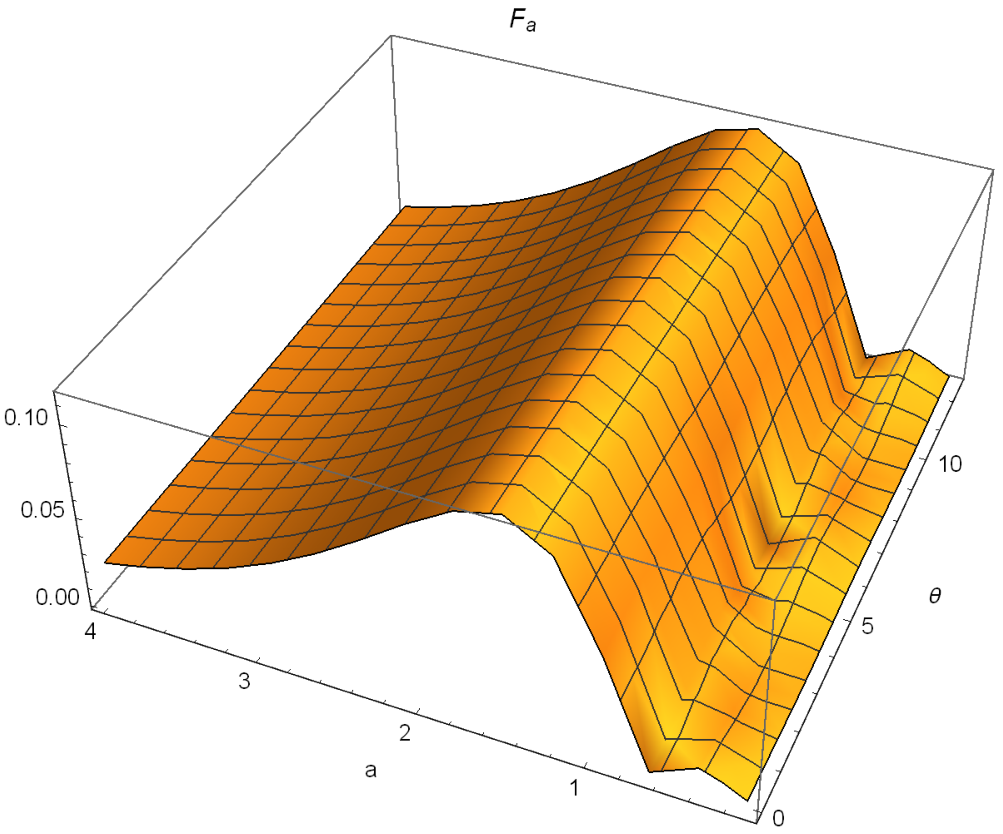}
\caption{\label{figzathF} The QFI of acceleration as a function of $a$ and $\theta$ for $z=0.5$. We take $\tau=5$ (left panel), $\tau=20$ (middle panel), and $\tau=40$ (right panel). }
\end{centering}
\end{figure}

We depict the QFI of the acceleration as a function of $a$ and $\theta$ for fixed $\tau$ in Fig. \ref{figzathF}. The periodicity of QFI with respect to the initial state parameter $\theta$ gradually vanishes with the evolution of time, which is similar to the unbounded case. However, the periodicity lasts a longer time than the previous unbounded case, which indicates that the initial state is protected by the boundary.

\subsection{Quantum estimation of temperature for a thermal bath with a boundary}

We consider a static atom immersed in a thermal bath with a boundary, and the field correlation function is given by
\begin{eqnarray}
G^+(t,t')=-\frac{1}{4\pi^2}\Sigma_{m=-\infty}^{\infty}\left[ \frac{1}{(t-t'-i m \beta-i \epsilon)^2}- \frac{1}{(t-t'-i m \beta-i \epsilon)^2-(2z)^2}\right].
\end{eqnarray}

The Fourier transformation of the field correlation function is given by
\begin{eqnarray}\label{fourierT}
{\cal G}(\lambda)=\frac{1}{2\pi}\frac{\lambda}{1-e^{-\lambda/T}}-\frac{1}{2\pi}\frac{\lambda}{1-e^{-\lambda/T}}\frac{\sin(2z \lambda)}{2z\lambda},
\end{eqnarray}
where $T={1}/{\beta}$.

The coefficients for the Kossakowski matrix are
\begin{eqnarray}\label{abT}
\begin{aligned}
&A_b=\frac{\mu^2 \omega_0 \coth\frac{\omega_0}{2T}}{8\pi}\left[1-\frac{\sin(2z\omega_0)}{2z\omega_0}\right]     \;,\\
&B_b=\frac{\mu^2 \omega_0 }{8\pi}\left[1-\frac{\sin(2z\omega_0)}{2z\omega_0}\right] \;.\\
\end{aligned}
\end{eqnarray}

In the following discussion, we use $t\rightarrow \tau\equiv{\mu^2\omega_0}t/{2\pi}$, $T\rightarrow \tilde{T}\equiv{T}/{\omega_0}$, and $z\rightarrow \tilde{z}\equiv z \omega_0$. For simplicity, $\tilde{T}$ and $\tilde{z}$ will be written as $T$ and $z$. We can obtain the QFI of the temperature $F_T$.

\begin{figure}[H]
\begin{centering}
\includegraphics[scale=0.52]{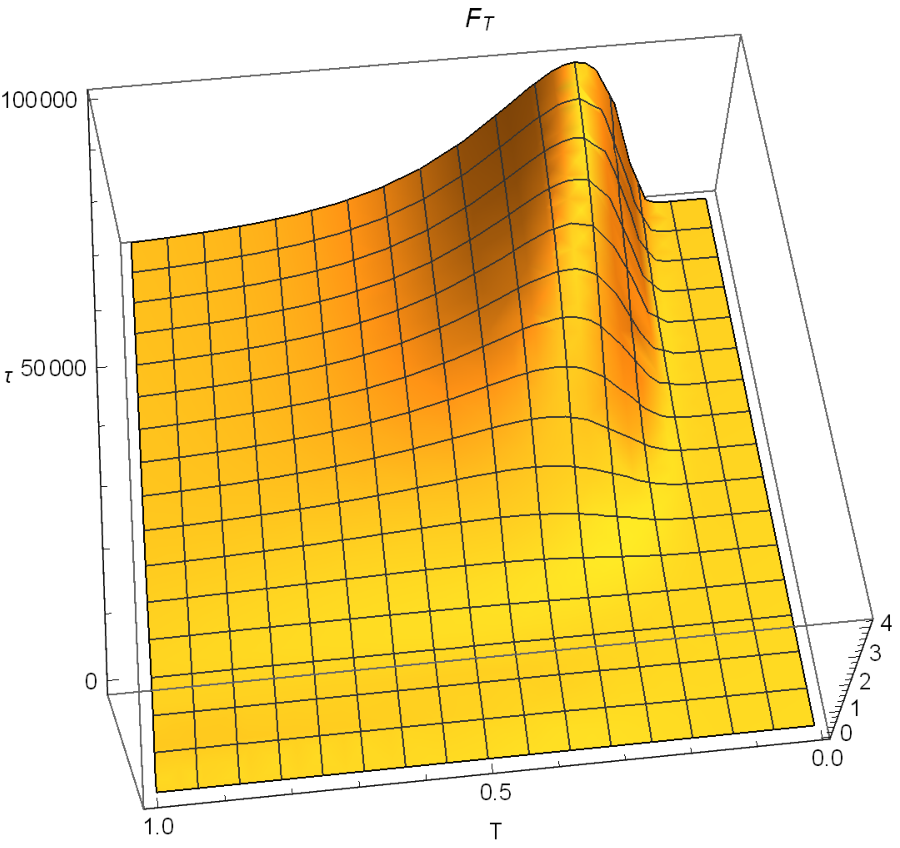}
\includegraphics[scale=0.52]{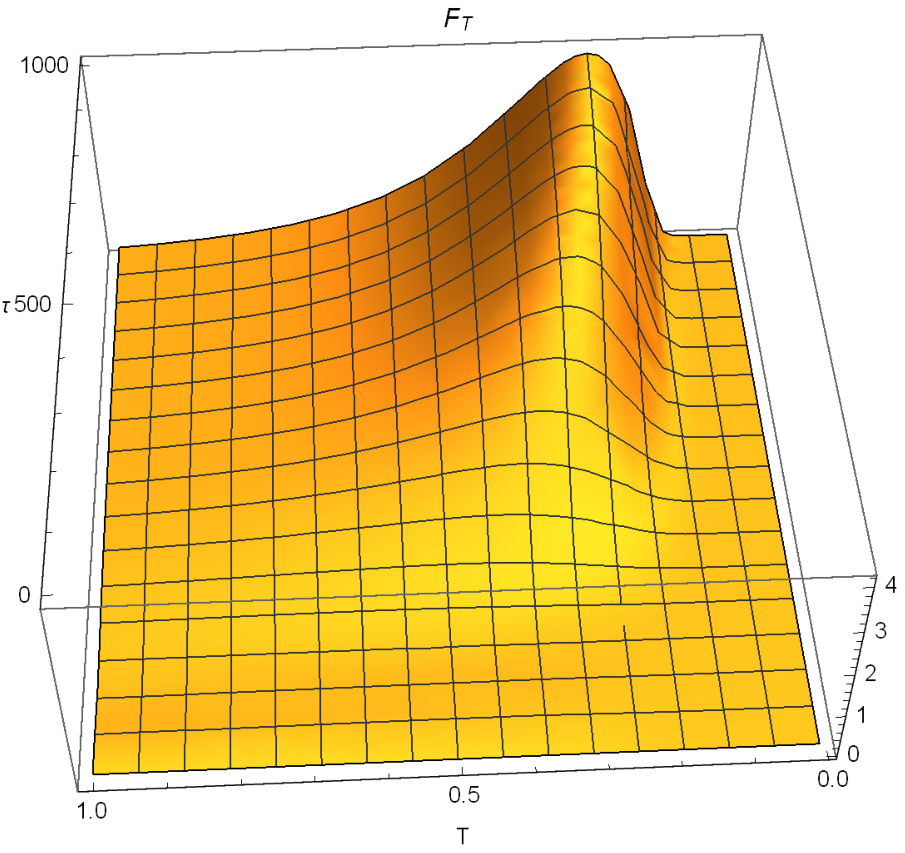}
\includegraphics[scale=0.53]{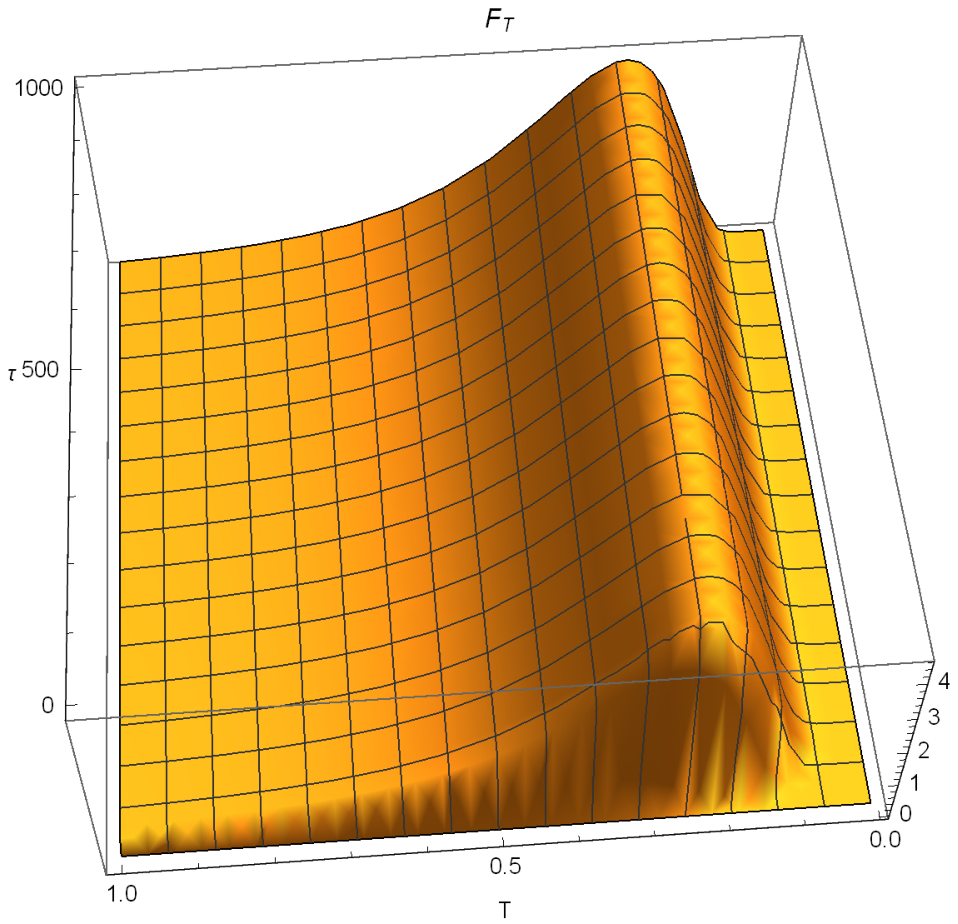}
\caption{\label{figTtF} The QFI of temperature as a function of $T$ and $\tau$ for $\theta=0$. We take $z=0.01$ (left panel), $z=0.5$ (middle panel), and $z=1$ (right panel). }
\end{centering}
\end{figure}

We describe the QFI of temperature as a function of $T$ and $\tau$ for fixed $\theta=0$ in Fig. \ref{figTtF}. Added a boundary, $F_{T}$ has no more than one peak value with a certain moment, which is similar to the unbounded case but different from the case of $F_{a}$. The QFI for $z=0.01$ is smaller than the case $ z=0.5$ and $ z=1$ in a short period, which is completely different from the case of a uniformly accelerated atom with a boundary.

\begin{figure}[H]
\begin{centering}
\includegraphics[scale=0.55]{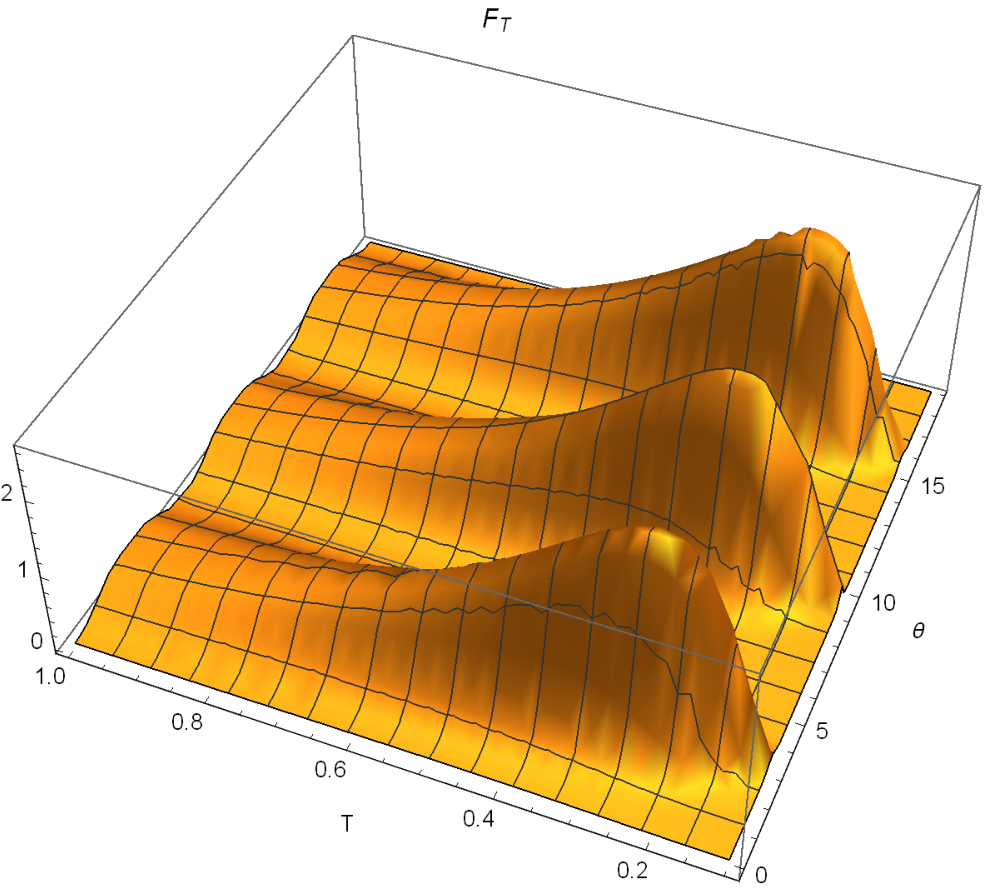}
\includegraphics[scale=0.55]{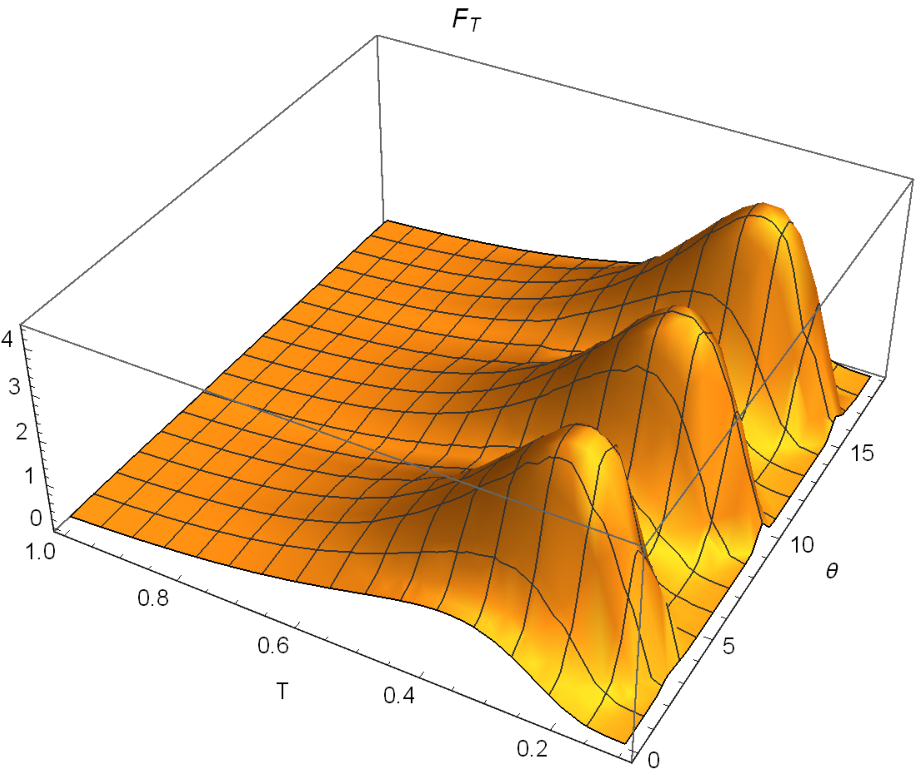}
\includegraphics[scale=0.55]{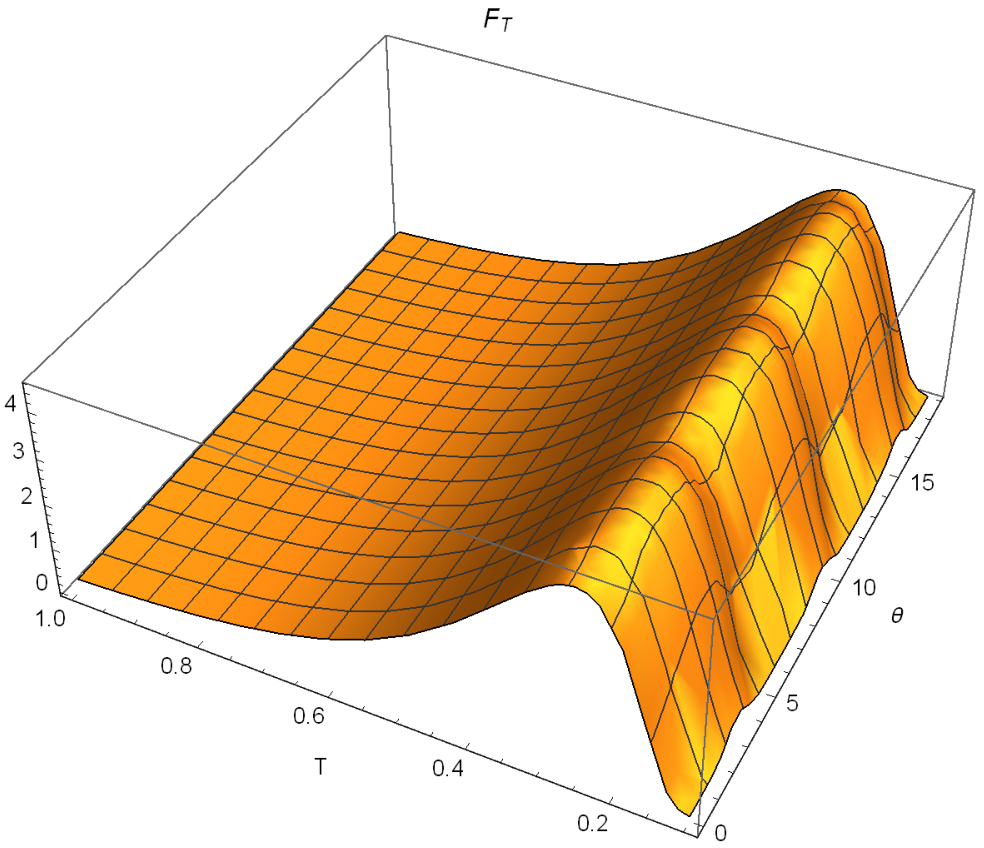}
\caption{\label{figTthF} The QFI of temperature as a function of $T$ and $\theta$ for $z=0.5$. We take $\tau=5$ (left panel), $\tau=20$ (middle panel), and $\tau=40$ (right panel). }
\end{centering}
\end{figure}

We depict the QFI of temperature as a function of $T$ and $\theta$ for fixed $z$ in Fig. \ref{figTthF}. The periodicity lasts a longer time than unbounded case, which shows the similarity in the case of a uniformly accelerated atom with a boundary. Therefore, we can conclude that the initial state is protected by the boundary.

\section{conclusion}

We have investigated, in an open quantum system, the evolution of a uniformly accelerated two-level atom which interacts with a massless scalar field in the Minkowski vacuum.
When the uniformly accelerated two-level detector evolves for a certain time, we found that the QFI initially increases and then decreases with increase of acceleration $a$, where there exists a peak value, which indicates that the optimal precision of estimation is achieved with an appropriate acceleration. The peak achieves faster for the ground state of the atom. It is shown that the QFI behaves periodically with respect to the initial state parameter $\theta$. However, the periodicity gradually fades away with the evolution of time. Therefore, we can deduce that the initial state cannot affect the later stable quantum state.

Adding a boundary, we found that the peak value of $F_a$ increases when the atom is close to the boundary, compared with the absence of boundary. It is obvious that the QFI is protected by the boundary, which means the estimation precision of acceleration is enhanced by adding a boundary. We observed that $F_a$ fluctuates, and there may exist two peak values with a certain moment, which expands the detection range of the acceleration. The periodicity of QFI with respect to the initial state parameter $\theta$ gradually vanishes with the evolution of time, which is similar to the unbounded case. However, the periodicity lasts a longer time than the previous unbounded case, which indicates that the initial state is protected by the boundary.

For a thermal bath with a boundary, $F_T$ has no more than one peak value with a certain moment, which is completely different from the case of a uniformly accelerated atom with a boundary. The periodicity lasts a longer time than unbounded case, which shows the similarity in the case of a uniformly accelerated atom with a boundary. Therefore, we can conclude that the initial state is protected by the boundary in the two cases.

\begin{acknowledgments}

This work was supported by the National Natural Science Foundation of China under Grants No. 11705144, No. 11775076, and No. 11875025, and the Scientific Research Program of Education Department of Shaanxi Provincial Government (Grant No. 17JK0706).

\end{acknowledgments}


\begin{thebibliography}{99}

\bibitem{Fulling} S. A. Fulling, Phys. Rev. D {\bf 7}, 2850 (1973).
\bibitem{Hawking1}S. W. Hawking, Nature(London) {\bf 248}, 30 (1974).
\bibitem{Hawking2}S. W. Hawking, Commun. Math. Phys. {\bf 43}, 199 (1975)
\bibitem{Davies1} P. C. W. Davies, J. Phys. A {\bf 8}, 609 (1975)
\bibitem{Unruh} W. G. Unruh, Phys. Rev. D {\bf 14}, 870 (1976).
\bibitem{Crispino} L. C. B. Crispino, A. Higuchi, and G. E. A. Matsas, Rev. Mod. Phys. {\bf80},787 (2008).
\bibitem{Lima} C. A. Uliana Lima, F. Brito, J. A. Hoyos, and D. A. Turolla Vanzella, Nat. Commun. {\bf10}, 3030(2019).
\bibitem{Bell} J. S. Bell and J. M. Leinaas, Nucl. Phys. B {\bf 212}, 131 (1983); Nucl. Phys. B {\bf 284}, 488 (1987).
\bibitem{Yablonovitch}E. Yablonovitch, Phys. Rev. Lett. {\bf 62}, 1742 (1989) .
\bibitem{Davies2} P. C. W. Davies,  Chaos {\bf 11}, 539 (2001).
\bibitem{Vanzella} D. A. T. Vanzella and G. E. A. Matsas, Phys. Rev. Lett. {\bf 87}, 151301 (2001).
\bibitem{Gibbons}  G. W. Gibbons and E. P. S. Shellard, Science {\bf 295}, 1476 (2002).
\bibitem{Milburn}P. M. Alsing and G. J. Milburn, Phys. Rev. Lett. {\bf 91}, 180404 (2003).
\bibitem{Fuentes} I. Fuentes-Schuller and R. B. Mann, Phys. Rev. Lett. {\bf 95}, 120404 (2005).
\bibitem{Helstrom} C. W. Helstrom, \emph{Quantum Detection and Estimation Theory}
(Academic, New York, 1976).
\bibitem{Holevo} A. S. Holevo, \emph{Probabilistic and Statistical Aspects of Quantum
Theory} (North-Holland, Amsterdam, 1982).
\bibitem{Cramer} H. Cram\'{e}r, \emph{Mathematical Methods of Statistics} (Princeton University, Princeton, NJ, 1946).
\bibitem{V. Buzek} V. Buzek, R. Derka, and S. Massar, Phys. Rev. Lett. {\bf 82}, 2207 (1999).
\bibitem{N. Poli} N. Poli, F.-Y. Wang, M. G. Tarallo, A. Alberti, M. Prevedelli, and G. M. Tino, Phys. Rev. Lett {\bf 106},038501 (2011).
\bibitem{Li} N. Li and S. Luo Phys. Rev. A {\bf 88}, 014301 (2013).
\bibitem{Giovannetti1} V. Giovannetti, S. Lloyd, and L. Maccone, Science {\bf 306}, 1330 (2004).
\bibitem{Giovannetti2} V. Giovannetti, S. Lloyd, and L. Maccone, Phys. Rev. Lett. {\bf 96}, 010401 (2006).
\bibitem{Giovannetti3} V. Giovannetti, S. Lloyd, and L. Maccone, Nat. Photonics. {\bf 5}, 222 (2011).
\bibitem{Sun}Y. Yao, X. Xiao, Li Ge, X.G. Wang, and C. P. Sun, Phys. Rev. A {\bf 89}, 042336 (2014).
\bibitem{Yu}Y. Jin and H. Yu, Phys. Rev. A. {\bf 91}, 022120 (2015).
\bibitem{Rajabpour} M. A. Rajabpour, Phys. Rev. D {\bf 96}, 126007 (2017).
\bibitem{Gessner} M. Gessner and A. Smerzi, Phys. Rev. A {\bf 97}, 022109 (2018).
\bibitem{Frowis}F. Frowis, M. Fadel, P. Treutlein, N. Gisin, and N. Brunner, Phys. Rev. A {\bf 99},040101(R) (2019).
\bibitem{Braunstein} S. L. Braunstein and C. M. Caves, Phys. Rev. Lett. {\bf 72}, 3439(1994); S. L. Braunstein, C. M. Caves, and G. J. Milburn, Ann. Phys. (N.Y.) {\bf 247}, 135 (1996).
\bibitem{Lindblad} V. Gorini, A. Kossakowski, and E. C. G. Surdarshan, J. Math. Phys. (N.Y.) {\bf 17}, 821 (1976); G. Lindblad, Commun. Math. Phys. {\bf 48}, 119 (1976).
\bibitem{pr5}F. Benatti, R. Floreanini, and M. Piani, Phys. Rev. Lett. {\bf 91}, 070402 (2003).
\bibitem{Benatti2} F. Benatti and  R. Floreanini, J. Opt. B {\bf 7}, S429 (2005).
\end{thebibliography}
\end{document}